\documentclass[aps,showpacs,nofootinbib,tightenlines]{revtex4}
\usepackage[colorlinks,urlcolor=blue,linkcolor=blue,citecolor=blue]{hyperref} %
\usepackage{amsmath,amssymb,braket,color,epsfig,graphicx,slashed,times}
\newcommand{\beq}{\begin{eqnarray}}
\newcommand{\eeq}{\end{eqnarray}}
\newcommand{\non}{\nonumber\\ }

\def \cpc{ Chin. Phys. C  }

\def \epjc{ Eur. Phys. J. C }

\def \jpg{  J. Phys. G }
\def \npb{  Nucl. Phys. B }
\def \plb{  Phys. Lett. B }

\def \prd{  Phys. Rev. D }

\def \jhep{ JHEP }

\begin{document}

\title{Resonances $\phi(1020)$ and $\phi(1680)$ contributions for the three-body decays $B^+ \to D_s^+K\bar K$}

\author{Ai-Jun Ma}  \email{theoma@163.com}

\affiliation{Department of Mathematics and Physics,  Nanjing Institute of Technology, Nanjing, Jiangsu 211167, P.R. China}

\date{\today}
\begin{abstract}
We study the resonances $\phi(1020)$ and $\phi(1680)$ contributions for the three-body decays $B^+ \to D_s^+K\bar K$ in the
perturbative QCD approach.  The branching ratios for $B^+ \to D_s^+\phi(1020) \to D_s^+ K^+K^-$ and
$B^+ \to D_s^+\phi(1020) \to D_s^+ K^0\bar{K}^0$ are predicted to be $(1.53\pm0.23) \times 10^{-7}$
and $(1.02^{+0.19}_{-0.13} )\times 10^{-7}$, respectively.
The decay $B^+ \to D_s^+\phi(1680)$ with $\phi(1680)$ decays into $K^+K^-$ or  $K^0\bar{K}^0$,  has the branching fraction
$(6.94^{+1.83}_{-2.02})\times 10^{-9}$, which is about $5\%$ of the result for
$B^+ \to D_s^+\phi(1020) \to D_s^+ K^+K^-$.
\end{abstract}

\maketitle


The rare decay $B^+ \to D_s^+\phi(1020)$ presents a very clean channel for us to test the annihilation contribution in the
Standard Model (SM). This decay process has been extensively studied on theoretical and experimental sides during the past
decades, with the predictions for its branching fraction in the range of $1.30\times10^{-7}$-$1.88\times10^{-6}$ in the
SM~\cite{epjc24-121,plb540-241,prd76-057701,jpg37-015002,prd92-094016}. In addition, the small branching ratio makes this
process probably sensitive to the parameters of the physics beyond SM and its direct $CP$ violation which is expected to be zero
in SM could also be produced in the new physics models~\cite{plb540-241,prd76-057701}. The search for $B^+ \to D_s^+\phi(1020)$
was performed by CLEO~\cite{plb319-365} and BABAR Collaborations~\cite{prd73-011103} years ago, but no significant signal has
been observed. The first evidence for this decay was found with greater than $3\sigma$ significance by LHCb Collaboration
with the measured branching fraction $(1.87^{+1.25}_{-0.73}\pm{0.19}\pm{0.32})\times 10^{-6}$~\cite{jhep02-043}.
Recently, in their work~\cite{jhep01-131},  LHCb set a limit as ${\cal B}(B^+ \to D_s^+\phi(1020))< 4.9(4.2)\times 10^{-7}$
at $95\% (90\%)$ confidence level in the analysis of the three-body decay $B^+ \to D_s^+K^+K^-$ for this two-body subprocess,
which is roughly one order smaller than the previous result in~\cite{jhep02-043}. One should note that the $\phi(1020)$ meson is usually
reconstructed within $K\bar K$ final states in the experimental analysis~\cite{jhep02-043,jhep01-131,jhep1708-037,prd98-071103,
prd85-054023,prd85-112010}, but treated as a stable particle in the aforementioned theoretical studies.

Three-body hadronic $B$ meson decays are much more complicated than the two-body cases partly because of the three-body
effects and rescattering effects~\cite{npb899-247,jhep1710-117,1512-09284} and also because of entangled resonant and nonresonant
contributions. The resonant contributions in the three-body decays are related to the low energy scalar, vector and tensor resonant
states, and could be isolated from the total decay amplitudes and studied in the quasi-two-body framework~\cite{plb763-29,
1605-03889,prd96-113003}. At the edge of the Dalitz plot~\cite{Dalitz-plot}, the three final state particles are quasi-aligned in the
rest frame of the $B$ meson, while two of them move collinearly and recoil against the third meson. The factorization for the two-body
decays is still valid for this part of the phase space. Then the relevant decay processes can be represented as
$B\to R h_3\to h_1h_2h_3$ where $h_3$ represents the bachelor particle moves in the opposite direction and the $h_1h_2$ pair
proceeds by the intermediate resonance $R$. The studies on a series of charmless three-body hadronic $B$ meson decays have
been accomplished based  on the QCD factorization
(QCDF)~\cite{plb622-207,prd74-114009,prd79-094005,plb699-102,prd72-094003,prd76-094006,prd88-114014,prd89-074025,
prd94-094015,prd89-094007,epjc75-536,prd99-076010,2007-02558} and the perturbative QCD (PQCD) approach~\cite{plb561-258,
prd70-054006,prd89-074031,plb763-29,prd95-056008,prd96-036014,prd98-056019,epjc79-37,epjc80-394,epjc80-517,
jhep2003-162,prd101-111901,2006-08223}.

In the previous works~\cite{npb923-54,cpc43-073103,prd96-093011,plb788-468,plb791-342,epjc79-539,prd100-014017}, the $S$- and $P$-wave
$\pi\pi$, $K\pi$ and $D\pi$ resonance contributions to several three-body  $B\to Dh_1h_2$ decays have been studied within the PQCD
approach, and most of the theoretical predictions are in good agreement with the available experimental results.
In this work, we shall study  the contributions from the subprocesses $\phi(1020,1680) \to K^+K^-$ and
$\phi(1020,1680) \to K^0\bar{K}^0$ to the three-body decay $B^+ \to D_s^+K\bar K$ within PQCD approach.
In our framework for the quasi-two-body decays, the two-meson distribution amplitudes are introduced to
describe the interactions between the meson pair associated with the resonance. The relevant decay amplitude $\cal A$ for the
quasi-two-body decays $B \to  D  R \to D  h_1 h_2$ concerned in this work can be expressed as~\cite{plb561-258,prd70-054006}
\begin{eqnarray}
{\cal A}=\Phi_B\otimes H\otimes \Phi_{D}\otimes\Phi_{h_1h_2},
\end{eqnarray}
where $H$ is the hard kernel and  $\Phi_B$
($\Phi_{D}$,$\Phi_{h_1h_2}$) represents the $B$ meson ($D$ meson, $h_1h_2$ pair) distribution amplitude.

In the rest frame of $B$ meson, we could define the momenta of the $B$ meson, the kaon pair which is generated from the
intermediate states $\phi(1020,1680)$, and the $D$ meson in the light-cone coordinates as
\beq\label{lc}
p_{B}&=&\frac{m_{B}}{\sqrt2}(1,1,\textbf{0}_{\rm T}),
~\quad p=\frac{m_{B}}{\sqrt2}(1-r^2,\eta,\textbf{0}_{\rm T}),~\quad
p_3=\frac{m_{B}}{\sqrt2}(r^2,1-\eta,\textbf{0}_{\rm T}),
\eeq
where the mass ratio $r = m_{D}/m_{B}$ and $m_{B(D)}$ is the mass for $B(D)$ meson. The variable $\eta$ is defined as
$\eta=s/[(1-r^2)m^2_{B}]$ with  the invariant mass square $s=p^2=m^2_{K\bar K}$ for the kaon pair.
The momenta of the light quarks in the corresponding states are chosen as $k_B$, $k$ and $k_3$, respectively, with
\beq
k_B=(0,x_Bp_B^-,\textbf{k}_{B \rm T}),~\quad k=(zp^+,0,\textbf{k}_{\rm T}),~\quad
k_3=(0,x_3p_3^-,\textbf{k}_{3\rm T}),
\eeq
where the momentum fractions $x_{B}$, $z$ and $x_3$ run between zero and unity in the numerical calculation.

In this work, we adopt the same distribution amplitudes for the $B^+$ and $D_s^+$ mesons as those in
Ref.~\cite{npb923-54,prd96-093011,prd86-114025}. The $P$-wave $K\bar K$ system distribution amplitudes are organized
into~\cite{prd101-111901,2006-08223}
\beq
  \phi^{P\text{-wave}}_{K\bar K}(z,s)=\frac{-1}{\sqrt{2N_c}}
      \left[\sqrt{s}\,{\epsilon\hspace{-1.5truemm}/}\!_L\phi^0(z,s) + {\epsilon\hspace{-1.5truemm}/}\!_L {p\hspace{-1.7truemm}/} \phi^t(z,s)
             +\sqrt s \phi^s(z,s)  \right],
\eeq
with the distribution amplitudes
\beq
   \phi^{0}(z,s)&=&\frac{3F_K(s)}{\sqrt{2N_c}} z(1-z)\left[1+a_2^{\phi} C^{3/2}_2(1-2z) \right],\label{def-DA-0}\\
   \phi^{s}(z,s)&=&\frac{3F^s_K(s)}{2\sqrt{2N_c}}(1-2z), \label{def-DA-s}\\    
   \phi^{t}(z,s)&=&\frac{3F^t_K(s)}{2\sqrt{2N_c}}(1-2z)^2.\label{def-DA-t}  
\eeq
The Gegenbauer polynomial $C^{3/2}_2(t)=3\left(5t^2-1\right)/2$ and the Gegenbauer moment
$a_2^{\phi}=0.18\pm0.08$ is the same as employed in Ref.~\cite{prd76-074018} for the two-body $B$ decays.
When concern only the resonance contributions, the relation between the kaon time-like form factors $F^{K^+K^-}_{s}$,
$F^{K^0\bar K^0}_{s}$ and the kaon electromagnetic form factors $F_\phi^{KK}$ can be written as~\cite{prd88-114014,2006-08223}
\beq
F_K(s)=F^{K^+K^-}_{s}(s)=F^{K^0\bar K^0}_{s}(s)=-3 F_\phi^{KK}(s)=-\sum_\phi c^K_\phi {\rm BW}_\phi(s).
\eeq
For the $F^{s,t}_K(s)$ in the distribution amplitudes, we adopt the relation
$F^{s,t}_K(s)\approx (f^T_{\phi}/f_{\phi})F_K(s)$~\cite{2006-08223} with the ratio $f^T_{\phi}/f_{\phi}=0.75$ at the scale
$\mu=2 {\rm~GeV}$~\cite{prd78-114509}.
The parameters $c^K_\phi$ have been fitted to the data in Refs.~\cite{epjc39-41,prd81-094014,jetp129-386},
we adopt the values $c^K_\phi(1020)=1.038$ and $c^K_{\phi(1680)}=-0.150\pm0.009$~\cite{jetp129-386} as those are discussed and   chosen in Ref.~\cite{2006-08223}.

\begin{figure}[tbp]
\begin{center}
\vspace{-2cm}
\centerline{\epsfxsize=16cm \epsffile{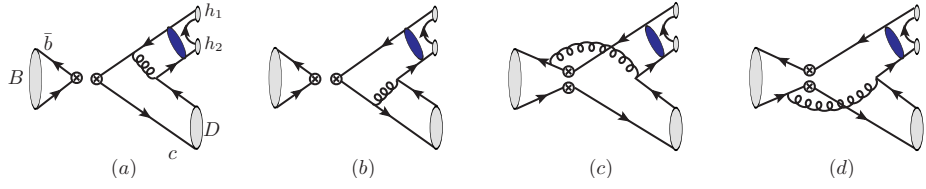}}
\vspace{-18cm}
\caption{Typical Feynman diagrams for the quasi-two-body decays $B^+ \to D_s^+ \phi(1020,1680) \to D_s^+ K\bar K$.
The $h_1h_2$ is the kaon pair and the ellipses represent the intermediate states $\phi(1020)$ and $\phi(1680)$.
  }
  \vspace{-1cm}
\label{fig:fig1}
\end{center}
\end{figure}

There are only tree operators contribute to the decay amplitude of the decays $B^+ \to {D_s^+} \phi(1020,1680) \to  {D_s^+} K\bar K$,
which can be written as
\beq
\mathcal{A}({B^+ \to {D_s^+} \phi \to  {D_s^+} K\bar{K} })&=&
\frac{G_F}{\sqrt{2}}V^*_{ub}V_{cs} [(\frac{C_1}{3}+C_2)F_{aD}^{LL}+C_1M_{aD}^{LL} ],
\eeq
where $G_F$ represents the Fermi coupling constant, $V_{ij}$ are the CKM matrix elements and $C_{1,2}$ mean the Wilson coefficients.
The symbols $F_{aD}^{LL}$ and $M_{aD}^{LL}$ are the amplitudes from the factorizable and nonfactorizable annihilation diagrams
shown in Fig.~\ref{fig:fig1}, respectively, with the specific expressions given by
\beq
\label{FaD}
 F_{aD}^{LL}&=&8\pi C_F m^4_B f_B \int dx_3 dz \int b_3 db_3 b db \phi_D\non
&&\times \big\{\big[  (r^2-1)[\eta(\eta+r^2-1)-(1-\eta)^2 x_3]\phi_0+2 r\sqrt{\eta (1-r^2) }[  1+\eta+(1-\eta)x_3-r^2] \phi_s \big]\non
&& \times E_a(t_a)h_a(z,x_3,b_3,b)S_t(x_3)-\big[[ (1-\eta)(r^4(z-1)+r^2(1-\eta-2z)+z-2rr_c)+2r^3r_c] \phi_0\non
&&+\sqrt{\eta (1-r^2) } [r(2z+2r^2(1-z)-rr_c)(\phi_s+\phi_t)+(1-\eta)(2r-r_c)(\phi_s-\phi_t)]\big]\non
&&\times E_a(t_b)h_b(z,x_3,b,b_3)S_t(z) \big\},
\eeq
\beq
\label{MaD}
 M_{aD}^{LL}&=&32\pi C_F m^4_B/\sqrt{6}  \int dx_B dz dx_3 \int b_B db_B b db \phi_B\phi_D\non
&&\times \big\{\big[(\eta+r^2-1)[(1-\eta)(r^2(z-x_3)-x_B-z)+r^2-\eta] \phi_0+r\sqrt{\eta (1-r^2) }[  (z(1-r^2)+x_B)\non
&&\times(\phi_s+\phi_t)+(1-\eta)x_3(\phi_s-\phi_t)+2\phi_s]  \big] E_n(t_c)h_c(x_B,z,x_3,b,b_B)\non
&&-\big[   (1-\eta +r^2 ) [(1-r^2)((1-\eta)x_3-\eta z)+x_B\eta ] \phi_0+r\sqrt{\eta (1-r^2) }\big[ (1-\eta)x_3(\phi_s+\phi_t)\non
&&+((1-r^2)z-x_B)(\phi_s-\phi_t)]\big] E_n(t_d)h_d(x_B,z,x_3,b,b_B)
\big\},
\eeq
with the color factor $C_F=\frac{4}{3}$. The explicit expressions of the hard functions $h_i$, the evolution
factors $E(t_i)$ and the threshold resummation factor $S_t$  can be found in Ref.~\cite{npb923-54}.

In the numerical calculation, we adopt the following input parameters~\cite{PDG2018}, with the QCD scale, masses, decay constants
and full widths in units of GeV,
\beq
\Lambda^{(f=4)}_{ \overline{MS} }&=&0.25, \quad m_{B}=5.279, \quad m_{D_s}=1.968,
\quad m_{K^\pm}=0.494,\quad m_{K^0}=0.498, \non
m_b&=&4.8, \quad m_c=1.275, \quad f_B=0.189, \quad f_{D_s}=0.249, \quad
\tau_{B}=1.638\; {\rm ps},  \non
m_{\phi(1020)}&=&1.019, \quad \Gamma_{\phi(1020)}=0.00425, \quad m_{\phi(1680)}=1.680, \quad \Gamma_{\phi(1680)}=0.150.
\label{eq:inputs}
\eeq
For the Wolfenstein parameters $(A, \lambda,\bar{\rho},\bar{\eta})$ of the CKM mixing matrix, we use the values
$A=0.836\pm0.015,~\lambda=0.22453\pm0.00044$,~$\bar{\rho} = 0.122^{+0.018}_{-0.017},~\bar{\eta}= 0.355^{+0.012}_{-0.011}$~\cite{PDG2018}.

The differential branching fractions ($\mathcal B$) for the quasi-two-body decays $B\to D \phi(1020,1680) \to DK\bar K$
can be written as~\cite{prd79-094005,prd101-111901,2006-08223}
\beq
 \frac{d{\mathcal B}}{d\eta}=\tau_B\frac{q^3 q^3_D}{12\pi^3m^5_B}|{\mathcal A}|^2\;.
\label{eqn-diff-bra}
\eeq
The magnitudes of the momenta for $K$ and $D$ in the center-of-mass frame of the kaon pair are written as
\beq
       q&=&\frac{1}{2}\sqrt{s-4m^2_K},  \label{def-q}\\
   q_D&=&\frac{1}{2\sqrt s}\sqrt{\left(m^2_{B}-m_{D}^2\right)^2 -2\left(m^2_{B}+m_{D}^2\right)s+s^2}.   \label{def-qh}
\eeq

By employing the decay amplitudes as given in Eq.~(\ref{FaD})-(\ref{MaD}) and the differential branching fractions in
Eq.~(\ref{eqn-diff-bra}), integrating over the full $K\bar K$ invariant mass region  $2m_K\leq \sqrt{s} \leq (m_{B^+}-m_{D_s^+})$
 for the resonant components,
we obtain the branching ratios
\beq
{\cal B}(B^+ \to D_s^+\phi(1020) \to D_s^+ K^+K^-)&=&(1.53\pm0.17(\omega_B)^{+0.14}_{-0.12}(a_{2\phi})^{+0.07}_{-0.10}(C_{D_s}))
\times 10^{-7}, \non
{\cal B}(B^+ \to D_s^+\phi(1020) \to D_s^+ K^0\bar{K}^0)&=&(1.02^{+0.13}_{-0.09}(\omega_B)^{+0.12}_{-0.08}(a_{2\phi})^{+0.06}_{-0.05}(C_{D_s})
 )\times 10^{-7},
\eeq
where the first error comes from the uncertainty of the $B$ meson shape parameter $\omega_B = 0.40 \pm 0.04$ {\rm GeV},
the second error comes from the Gegenbauer coefficient $a_2^{\phi}=0.18\pm0.08$ in the kaon-kaon distribution amplitudes and
the last one is induced by $C_{D_s}= 0.4\pm 0.1$ for $D_s$ meson wave function. The errors come from the uncertainties of other
parameters are small and have been neglected. Under the narrow-width approximation, the two-body branching fraction for
$B \to D \phi(1020)$ can be extracted from the quasi-two-body prediction with the relation
\beq
{\cal B}(B \to  D  \phi(1020) \to D  K \bar{K}) \approx  {\cal B}(B \to  D  \phi(1020)) \cdot {\cal B}(\phi(1020) \to K \bar{K}),
\label{qs2}
\eeq
Utilizing  the decay rate ${\cal B}(\phi(1020) \to K^+K^-)=0.492$~\cite{PDG2018}, we have the two-body branching fraction
${\cal B}(B^+ \to D_s^+\phi(1020) )=(3.11\pm0.47)\times10^{-7}$. The corresponding experimental results are given as
\beq  
{\cal B}(B^+ \to D_s^+\phi(1020))\left\{\begin{array}{ll}
     < 3\times 10^{-4}                                        {\hspace{4cm}}  {\rm CLEO}~$\cite{plb319-365}$,  \\
     < 1.9\times 10^{-6}                                     {\hspace{3.5cm}}~~ {\rm BABAR}~$\cite{prd73-011103}$,  \\
     =(1.87^{+1.25}_{-0.73}\pm{0.19}\pm{0.32})\times 10^{-6}  \quad~ {\rm LHCb}~$\cite{jhep02-043}$,  \\
     =(1.2^{+1.6}_{-1.4}\pm{0.8}\pm{0.1})\times 10^{-7}            {\hspace{1cm}}~{\rm LHCb}~$\cite{jhep01-131}$.
\end{array} \right.
\eeq
The branching fraction predicted in this work is consistents with the experiment data and limits.
The branching fraction for the two-body decay $B^+ \to D_s^+\phi(1020)$ has been calculated in~\cite{epjc24-121,jpg37-015002}
within PQCD approach, with the result consistents with our prediction within errors.
Comparing with the relatively large branching ratios predicted by other works~\cite{plb540-241,prd76-057701,prd92-094016},
the measured result by LHCb in~\cite{jhep01-131} is more closer to the PQCD prediction in this work.

The ratio between the branching fractions of the decays $B^+ \to D_s^+\phi(1020) \to D_s^+ K^0\bar{K}^0$
and $B^+ \to D_s^+\phi(1020) \to D_s^+ K^+K^-$ is defined as
\beq
 R_1=\frac{{\cal B}(B^+ \to D_s^+\phi(1020) \to D_s^+ K^0\bar{K}^0)}{{\cal B}(B^+ \to D_s^+\phi(1020) \to D_s^+ K^+K^-)}
        \approx 0.67.
\eeq
Based on the Eq.~(\ref{qs2}), we have
\beq
R_1 \approx \frac{{\cal B}(\phi(1020) \to K^0\bar{K}^0)}{{\cal B}(\phi(1020) \to K^+K^-)}.
\eeq
Then we estimate ${\cal B}(\phi(1020) \to K^0\bar{K}^0)=0.33$ with ${\cal B}(\phi(1020) \to K^+K^-)=0.492$~\cite{PDG2018},
which is agree with ${\cal B}(\phi(1020) \to K_L^0K_S^0)=0.340\pm0.004$ in the {\it Review of Particle Physics}~\cite{PDG2018}.

The prediction for the branching ratio involves $\phi(1680)$ is
\beq
{\cal B}(B^+ \to D_s^+\phi(1680) \to D_s^+ K^+K^-)=(6.94^{+0.90}_{-0.81}(\omega_B)^{+1.29}_{-1.62}(a_{2\phi})^{+0.35}_{-0.21}(C_{D_s})
\pm0.86(c_\phi^K)) \times 10^{-9},
\eeq
with the last error comes from the coefficient $c^K_{\phi(1680)}=-0.150\pm0.009$ in the form factor $F_K$.
Different from the decay modes with the subprocesses $\phi(1020)\to K^0\bar K^0$ and $\phi(1020)\to K^+ K^-$,
the decay $B^+ \to D_s^+\phi(1680) \to D_s^+ K^0\bar{K}^0$ almost has the same branching fraction as the decay
$B^+ \to D_s^+\phi(1680) \to D_s^+ K^+K^-$ because of the ratio
$\frac{{\cal B}(\phi(1680) \to K^0\bar{K}^0)}{{\cal B}(\phi(1680) \to K^+K^-)}\approx 1$~\cite{2006-08223}.
From another perspective, the main portion of the related branching ratios come from the region around the pole mass of the
resonant states, the lower limit of integration $2m_K$ is close to the pole mass of $\phi(1020)$ but relatively far away from that
 of $\phi(1680)$ which makes the branching ratios of the decay $B^+ \to D_s^+\phi(1020) \to D_s^+ K\bar{K}$ more
sensitive to the mass of kaon.
We can define the ratio $R_2$ between the branching fractions for $\phi(1680) \to K^+K^-$ and $\phi(1020) \to K^+K^-$ as
\beq
 R_2\approx\frac{{\cal B}(\phi(1680) \to K^+K^-)}{{\cal B}(\phi(1020) \to K^+K^-)}
        \approx \frac{{\cal B}(B^+ \to D_s^+\phi(1680) \to D_s^+ K^+K^-)}{{\cal B}(B^+ \to D_s^+\phi(1020) \to D_s^+ K^+K^-)}
        \approx0.05,
\eeq
which is consistent with the result $0.06$ obtained from the fit fractions $(70.5\pm0.6\pm1.2)\%$ and $(4.0\pm0.3\pm0.3)\%$
for the contributions of $\phi(1020)$ and $\phi(1680)$ in $B^0_s\to J/\psi K^+K^-$ decay~\cite{jhep1708-037}.
With the branching ratio $ {\cal B}{(B^+ \to  D_s^+ K^+K^-)}=(7.1\pm0.5\pm0.6\pm0.7)\times 10^{-6}$ presented by
LHCb~\cite{jhep01-131}, one has the percent at about $2.2\%$ of the total branching fraction  for the quasi-two-body decay $B^+ \to D_s^+\phi(1020) \to D_s^+ K^+K^-$,
which is expected to be tested in the future experiments.

To sum up, we studied the contributions for the $K^+K^-$ and $K^0\bar K^0$ originated from the intermediate states $\phi(1020)$
and $\phi(1680)$ in the three-body decays $B^+ \to D_s^+K\bar K$. The branching ratios for
$B^+ \to D_s^+\phi(1020) \to D_s^+ K^+K^-$ and $B^+ \to D_s^+\phi(1020) \to D_s^+ K^0\bar{K}^0$ are predicted to be
$(1.53\pm0.17^{+0.14+0.07}_{-0.12-0.10}) \times 10^{-7}$ and $(1.02^{+0.13+0.12+0.06}_{-0.09-0.08-0.05} )\times 10^{-7}$,
respectively, in this work. The branching ratio extracted from the quasi-two-body result for the two-body decay
$B^+ \to D_s^+\phi(1020)$ agrees with the existing experiment data within errors.
The decay $B^+ \to D_s^+\phi(1680)$ with $\phi(1680)$ decays into $K^+K^-$ or  $K^0\bar{K}^0$,
has the branching fraction $(6.94^{+0.90+1.29+0.35}_{-0.81-1.62-0.21}\pm0.86)\times 10^{-9}$, which is about $5\%$ of the
result for $B^+ \to D_s^+\phi(1020) \to D_s^+ K^+K^-$.

\begin{acknowledgments}
 Many thanks to  Wen-Fei Wang for valuable discussions. This work was supported by the National Natural Science Foundation of China under the Grant No.~11947011,
 the Natural Science Foundation of Jiangsu Province under Grant No.~BK20191010 and the Scientific Research
 Foundation of Nanjing Institute of Technology under Grant No.~YKJ201854.
\end{acknowledgments}


\end{document}